\documentclass[letterpaper, 12pt]{ewshm}

\usepackage[dvips,final]{graphicx}

\usepackage[table]{xcolor}
\usepackage{makecell}
\usepackage{enumerate}
\usepackage{color}
\usepackage{subfig}
\usepackage{caption}
\usepackage{amsfonts}
\usepackage{amsmath}
\usepackage{amssymb}
\usepackage{times}
\usepackage{cite}
\usepackage{hhline}
\usepackage{compactbib}
\usepackage{graphicx}        
\usepackage{amsmath,amsfonts,amssymb}
\usepackage{geometry}
\usepackage{lineno,hyperref}
\usepackage[labelsep=period]{caption}

\definecolor{halfgreen}{RGB}{0,128,0}
\definecolor{ahsred}{RGB}{192,0,0}

\renewcommand{\thefigure}{\arabic{figure}}
\renewcommand{\thetable}{\Roman{table}}

\newcommand{\beq}{\begin{equation}}
\newcommand{\eeq}{\end{equation}}
\newcommand{\bgqar}{\begin{eqnarray}}
\newcommand{\enqar}{\end{eqnarray}}
\newcommand{\bgqarn}{\begin{eqnarray*}}
\newcommand{\enqarn}{\end{eqnarray*}}
\newcommand{\bgary}{\begin{array}}
\newcommand{\enary}{\end{array}}

\long\def\symbolfootnote[#1]#2{\begingroup%
\def\thefootnote{\fnsymbol{footnote}}\footnote[#1]{#2}\endgroup}

\makeatletter
\renewcommand\@biblabel[1]{#1.}
\makeatother

\begin{document}


\vspace*{4.4cm}

\noindent Title: \textbf{Detecting Gait Abnormalities in Foot-Floor Contacts During Walking Through Footstep-Induced Structural Vibrations}

\vspace{1.6cm}

\noindent
$
\begin{array}{ll}
\text{Authors}: 
& \text{Yiwen Dong}  \\ 
& \text{Yuyan Wu} \\
& \text{Haeyoung Noh} 
\end{array}
$

\newpage


\vspace*{60mm}

\noindent \uppercase{\textbf{ABSTRACT}} \vspace{12pt} 

This study aims to detect abnormal human gait patterns through the dynamic response of floor structures during foot-floor interactions. Gait abnormality detection is critical for the early discovery and progressive tracking of musculoskeletal and neurological disorders, such as Parkinson's and Cerebral Palsy. Especially, analyzing the foot-floor contacts during walking provides important insights into gait patterns, such as contact area, contact force, and contact time, enabling gait abnormality detection through these measurements. Existing studies use various sensing devices to capture such information, including cameras, wearables, and force plates. However, the former two lack force-related information, making it difficult to identify the causes of gait health issues, while the latter has limited coverage of the walking path.

In this study, we leverage footstep-induced structural vibrations to infer foot-floor contact profiles, which allows force-informed and more wide-ranged gait abnormality detection. The main challenge lies in modeling the complex force transfer mechanism between the foot and the floor surfaces, leading to difficulty in reconstructing the force and contact profile during foot-floor interaction using structural vibrations. To overcome the challenge, we first characterize the floor vibration for each contact type (e.g., heel, midfoot, and toe contact) to understand how contact forces and areas affect the induced floor vibration. Then, we leverage the time-frequency response spectrum resulting from those contacts to develop features that are representative of each contact type. Finally, gait abnormalities are detected by comparing the predicted foot-floor contact force and motion with the healthy gait. To evaluate our approach, we conducted a real-world walking experiment with 8 subjects. Our approach achieves 91.6\% and 96.7\% accuracy in predicting contact type and time, respectively, leading to 91.9\% accuracy in detecting various types of gait abnormalities, including asymmetry, dragging, and midfoot/toe contacts. 


\symbolfootnote[0]{\hspace*{-7mm} Yiwen Dong, Ph.D. Candidate, Email: ywdong@stanford.edu. Structures as Sensors Lab, Department of Civil and Environmental Engineering, Stanford University, Stanford, CA, USA.}


\vspace{12pt} 
\noindent \uppercase{\textbf{INTRODUCTION}}  \vspace{12pt} 

Structural vibrations induced by human footsteps during walking contain important gait pattern information, allowing ubiquitous gait health monitoring and abnormality detection in daily life~\cite{dong2020md,dongstructure}. Detecting gait abnormalities is critical for the early discovery and progressive tracking of musculoskeletal and neurological disorders, such as Parkinson's and Cerebral Palsy~\cite{pistacchi2017gait,martin2006gait,deluca1991gait}. Gait abnormalities are typically reflected in deviations of posture, balance, and speed of walking from normal walking patterns, which indicates underlying conditions that affect the muscle groups or nervous systems~\cite{agostini2013segmentation}. Among them, analyzing the dynamic interaction between the foot and the floor during walking is a critical aspect because it helps to understand how the ground reaction forces act and transmit through the body~\cite{mickelborough2000validity}. Modeling the foot-floor contact can inform the gait health status to provide better design of medical interventions in neurological/musculoskeletal disorders, manage fall risks, and prevent further injuries or complications.

There are existing studies that detect gait abnormalities during walking, including force plates, wearable devices, and cameras~\cite{potluri2019deep,liu2011mobile,purcell2006use,nguyen2016skeleton,chen2008intelligent}. Force plates are the most commonly used technique for measuring foot-floor contact forces during walking. However, the coverage of a force plate is typically limited to one footstep only. Cameras and wearables measure the movement of the foot. They are more portable and wide-ranged than force plates and are more practical for everyday use. However, the camera measurements are limited to kinematic aspects of the gait, making it difficult to infer the joint forces and identify abnormal loading patterns. The wearables require a person to carry devices, which may cause discomfort and inconvenience. 

In this study, we leverage footstep-induced floor vibrations to detect gait abnormalities in foot-floor contacts during walking. The primary intuition is, as a person's foot contacts the floor and generates structural vibrations during walking, we capture those vibrations through floor-mounted geophone sensors. By analyzing the vibration signals, we infer the foot-floor contact profiles, including contact type and duration. Our approach is non-invasive, wide-ranged (up to 20 m range~\cite{pan2016occupant}), and produces more comprehensive gait information that informs both force and movement. 

The main challenge of this study is the complex mechanism during the foot-floor contact - not only the characteristics of the contact (i.e., force magnitude, direction, contact area) are changing when walking, they are also entangled with the dynamic property of the structure in the observed vibrations. To overcome this challenge, we first identify foot-floor contact types that are commonly described in medical settings and characterize their influence on the resultant vibrations. Then, we leverage the time-frequency response spectrum resulting from those contacts to develop features that are representative of each contact type. Finally, we detect abnormal gait patterns through machine learning, which translates the model estimations into clinical insights about gait health.

The contributions of the study are that we:
    \vspace{-0.1in}
\begin{enumerate}
    \item Develop a novel approach to detect abnormal gait using footstep-induced floor vibrations;
    \vspace{-0.1in}
    \item Characterize structural vibrations induced by foot-floor contacts to develop representative features; and
    \vspace{-0.1in}
    \item Evaluate our approach through a real-world human experiment to demonstrate its effectiveness
\end{enumerate}
    \vspace{-0.05in}

We conducted a real-world walking experiment with 8 human subjects. Our approach achieves 91.6\% and 96.7\% accuracy in predicting contact type and duration, respectively, and achieves a 91.9\% accuracy in detecting various types of gait abnormalities.  (including toe-walking, dragging, and asymmetry), which establishes its efficacy in gait health monitoring.


\vspace{20pt}

\noindent \uppercase{\textbf{Characterizing Footstep-induced Structural Vibrations for Abnormal Contact Detection}} \vspace{12pt}

 In this section, we characterize the complex mechanism during foot-floor contact when walking. Specifically, we first discuss how various contact types affect the resultant vibrations and then conduct time-frequency response analysis to quantify and interpret such effects.
 
\textbf{Background of Foot-Floor Contact Types}

\begin{figure}[!t] 
  \centering{\includegraphics[width=0.7\textwidth]{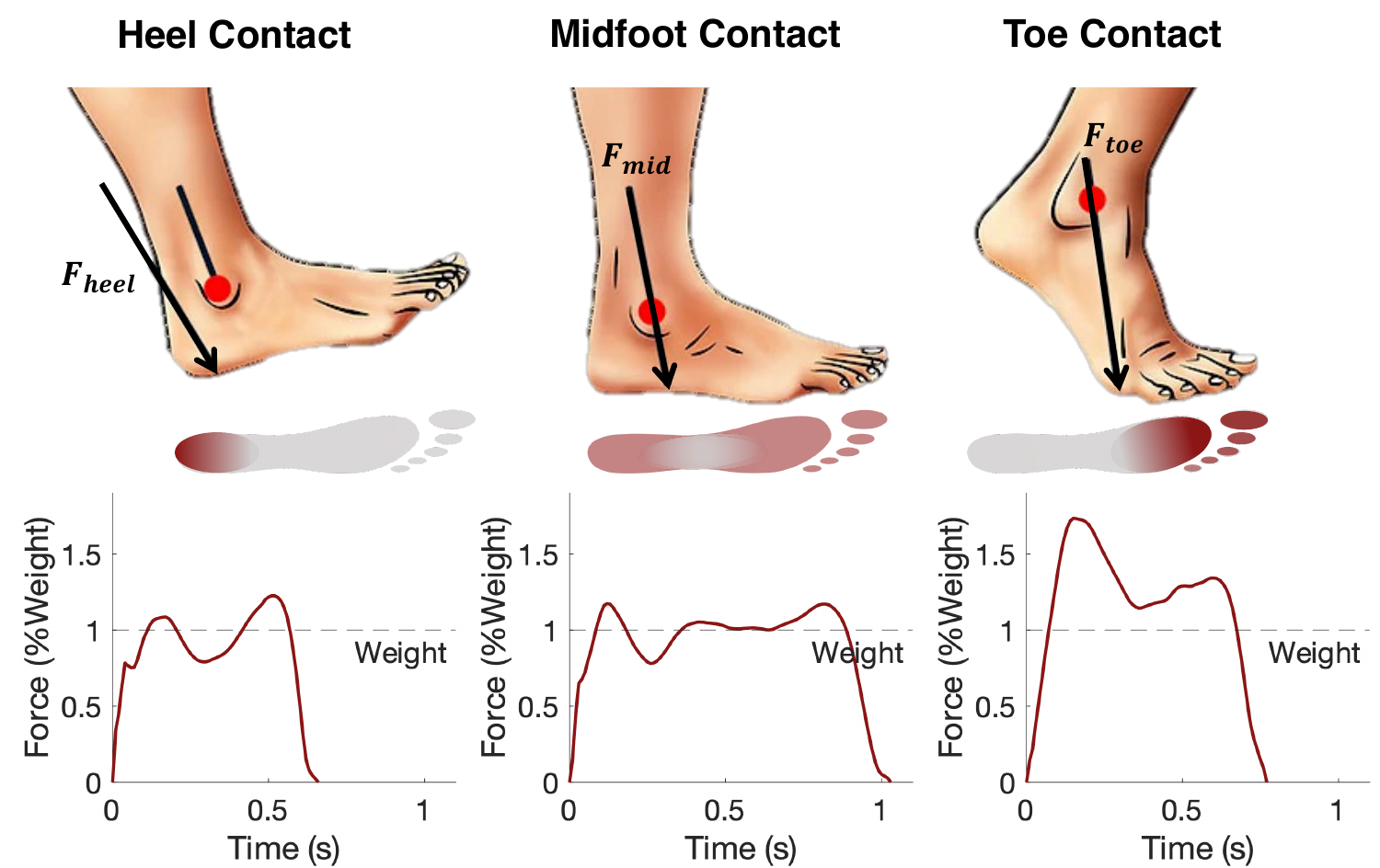}}
    \caption{Illustration of three types of foot-floor contacts and vertical footstep force variations during the contacts.}
\label{fig:contact_type}%
\end{figure}

To characterize various foot-floor contacts during walking, we first group them into three types that are commonly described in medical settings~\cite{buldt2018relationship,wyers2021foot}, including 1) heel contact, 2) midfoot contact, and 3) toe contact. As shown in Figure~\ref{fig:contact_type}, each type of contact has a distinct form in terms of contact area and force distribution. To understand the force transfer mechanism during the foot-floor contact, we also measure the footstep forces (ground reaction force (GRF) in clinical terms) during these contacts.

\begin{itemize}
    \item \textbf{Heel Contact:} A heel contact is when the heel of the foot strikes the floor first, which is typical in healthy adults. The force has both frictional and normal components and the normal component magnitude varies around the body weight.
    \item \textbf{Midfoot Contact:} A midfoot means the entire foot contacts the floor first, which is an indication of abnormal gait with insufficient ankle dorsiflexion. The duration of the contact is longer with fewer force variations.
    \item \textbf{Toe Contact:} A toe contact means the toe strikes the floor whereas the heel remains lifted. This is typically seen in individuals with later-stage gait disorders where excessive muscle contraction occurs. The force is significantly larger due to the impulse of the toe strike.
\end{itemize}


\textbf{Modeling Foot-Floor Contacts using Dynamic Floor Response Analysis}

\begin{figure}[!t] 
  \centering{\includegraphics[width=0.55\textwidth]{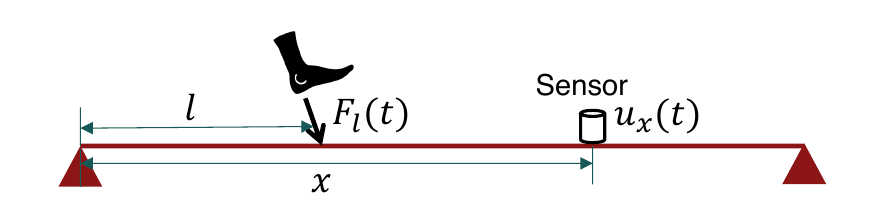}}
    \caption{Structural dynamics model for foot-floor contact.}
\label{fig:dynamicmodel}%
\end{figure}

To examine how various foot-floor contacts and structural properties affect the resultant floor vibration, we formulate the problem by simplifying the complex situation. We first assume that each footstep exerts a force $F_l(t)$ at a simply supported beam within the linear elastic range, as described in Figure~\ref{fig:dynamicmodel}. 

Based on the equation of motion~\cite{chopra2007dynamics}, we have:
\begin{equation}
    M\ddot{u}(t) + C\dot{u}(t) + Ku(t) = F_l(t) = -S(t)\ddot{u}_f(t)
\end{equation}\label{eq:1}
where $S(t)$ is the equivalent spatial loading matrix, which can be further expressed as the multiplication of body weight $M_{bw}$ and the spatial distribution of the force $\iota(t)$ (i.e., $S(t)=M_{bw}\iota(t)$). $\ddot{u}_f(t)$ is the foot acceleration, which is proportional to the force described in Figure~\ref{fig:contact_type}. 

In order to simplify the complex temporal and spatial dependency, we focus on the initial contact and assume the contact surface and force are constant during that time (i.e., $S(t) = S$). This is because the main difference between the three contact types is the initial contact force and surface, which results in distinct vibration patterns.  
Assuming temporal and spatial independence, we leverage modal decomposition and Fourier transform, resulting in the modal displacement response and the time-frequency response as follows:
\begin{equation}
    u_i(t) = P_i\phi_iD_i(t) \Rightarrow \mathcal{F}{u_i(t)} =  P_i \mathcal{F}{\phi_iD_i(t)}
\end{equation}\label{eq:2}
where $P_i = \frac{1}{m^*_i}\phi^T_iS$ is the modal participation factor of $i$-th mode; $\phi_i$ describes the mode shape; $D_i(t)$ describes the generalized modal displacement resulting from the foot acceleration $\ddot{u}_f(t)$.

In summary, the above derivation suggests that the \textit{time-frequency response of the floor} is determined by \textit{1) the body weight}, \textit{2) the spatial distribution of the footstep force}, \textit{3) the foot acceleration}, as in amplitude and direction during the contact. This formulation provides a theoretical foundation for modeling and predicting various foot-floor contact types.

\textbf{Time-Frequency Response Analysis for Typical Foot-Floor Contacts}

\begin{figure}[!t] 
  \centering{\includegraphics[width=0.75\textwidth]{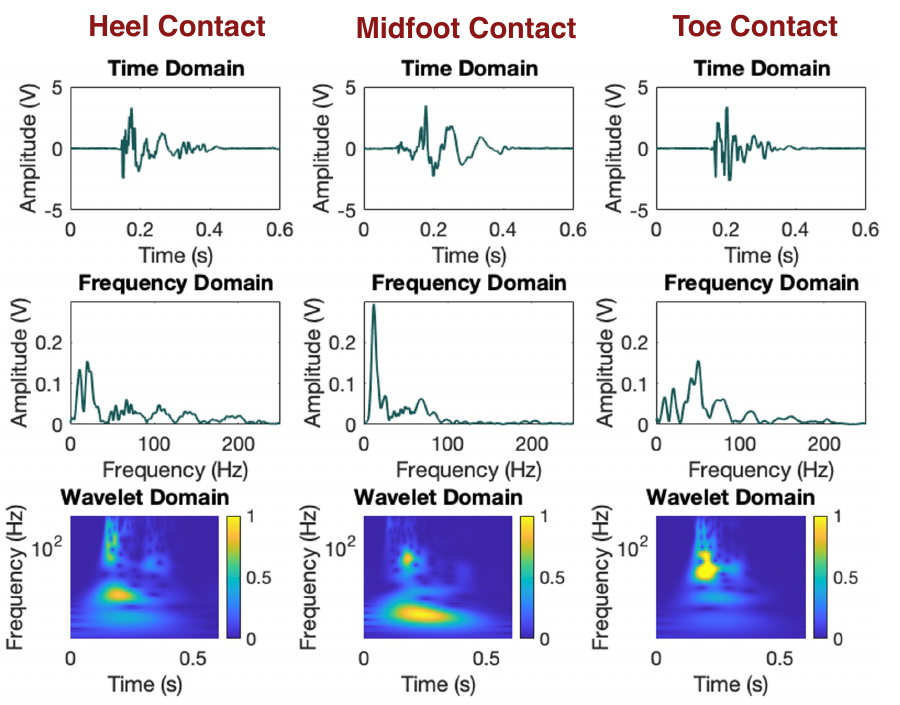}}
    \caption{Floor vibration induced by typical foot-floor contacts from time, frequency, and wavelet domain.}
\label{fig:vibration}%
\end{figure}

After establishing the theoretical basis, we conduct a time-frequency response analysis of the observed floor vibration to discuss its relation to typical contact characteristics. Figure~\ref{fig:vibration} demonstrates the time domain, frequency domain, and wavelet (time-frequency) domain data plots of the floor vibration under typical foot-floor contacts from the same person. The peaks and hotspots in the wavelet plot indicate the activated modes under a specific contact. We observe that midfoot contact induces lower modes (11, 53 Hz) than 
 heel/toe contacts (11, 53, 80, 112, 165, 200 Hz). 
 
To understand how the above observation is related to the foot-floor contact formulation, a qualitative analysis is summarized below:
\begin{itemize}
    \item \textbf{Effect of the Body Weight:} The amplitude in the frequency spectrum increases proportionally to the body weight because the modal participation factor is $S = M_{bw}\iota$. For the same person, the body weight does not affect the variation of the frequency spectrum across contact types as the body weight is typically constant. 
    \item \textbf{Effect of the Contact Area:} A larger contact area by midfoot contact typically induces lower modes of vibration than heel/toe contacts. This is because the modal participation factor is an inner product of the spatial loading matrix $S$ and mode shape $\phi_i$, which can be regarded as a weighted sum of the mode shape amplitudes within the contact area. For a wider spread load distribution $S$ for the midfoot contact, a higher mode produces a smaller sum because it is more likely to have oscillations in shape within the contact area that cancel out in the inner product. A lower mode is less likely to have oscillation within the contact area. On the other hand, a more concentrated load $S$ results in a higher inner product even for higher mode because it is less likely to have oscillations within that narrow contact area.
    \item \textbf{Effect of the Force Direction:} A force applied in the horizontal direction along the floor surface typically induces higher modes than the same force applied in the vertical direction. This is because the horizontal force affects the vertical vibration mainly through the floor thickness with waves that have shorter wavelengths. As a result, the modal participation factor for higher modes is larger for horizontal forces.
\end{itemize}
Since the area of midfoot contact is larger than the heel/toe contact and the force direction is mostly vertical, it tends to induce lower modes than the heel/toe contact. In addition, the toe contact has a less proportion of horizontal force than the heel contact, so it has a lower proportion of the high-frequency components ($\ge$100 Hz) in the spectrum.

Based on the analysis above, we develop features using the frequency spectrum of the vibration signals to represent the characteristics of the various foot-floor contact types. The features include the amplitudes at the union of dominant frequencies induced by these three contacts.

%


\vspace{20pt}
\noindent \uppercase{\textbf{Abnormal Gait Detection using Footstep-induced Floor Vibrations}} \vspace{12pt}

Based on clinical studies, abnormal foot-floor contacts typically manifest distinct contact types and durations. For example, toe-walking is a pathological gait with toe contacts, and limping has an asymmetrical contact duration~\cite{buldt2018relationship,deluca1991gait,wyers2021foot}. 

\begin{figure}[!t] 
  \centering{\includegraphics[width=0.75\textwidth]{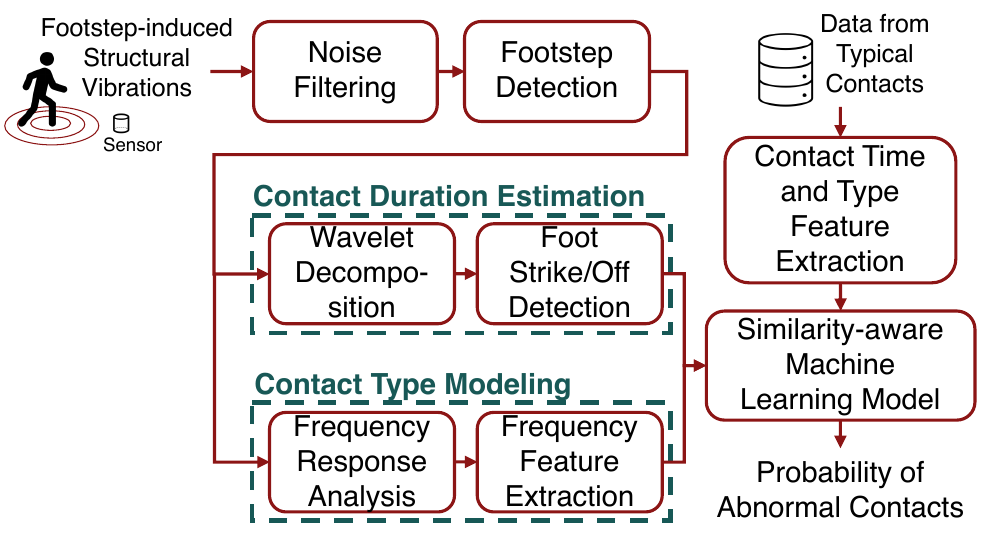}}
    \caption{The abnormal gait detection framework.}
\label{fig:system}%
\end{figure}

Our framework for abnormal gait detection incorporates both contact type and duration, shown in Figure~\ref{fig:system}. First, we pre-process the input floor vibrations generated from footsteps through noise filters and footstep detection algorithms - for noise filtering, we use the lowpass and wiener filters; for footstep detection, we set thresholds on cumulative wavelet coefficients~\cite{dong2020md}. Then, we detect the initial contact time when the higher frequencies are first activated and the foot-off time when the free vibration starts to dominate the spectrum, as developed in our prior work~\cite{dongstructure}. Meanwhile, we model the contact types using the frequency response features discussed in the previous section. Finally, we compare the similarity between these features and the pre-collected data from the typical contacts (both normal and abnormal) through a similarity-aware machine learning model to predict the probability of the incoming data being abnormal.

In our approach, a support vector machine with a radial basis function kernel is used to maximize the cluster margins between normal contact and each of the abnormal contact types. This is because the model has a good balance between bias and variance, and can learn from a limited amount of training data. 

To further empower the model to be aware of similarity, we apply Platt scaling to fit the margins for each sample, which provides a unified measurement of how similar the observed footsteps are to the abnormal ones~\cite{platt1999probabilistic}, described as follows:
\begin{equation}
    P(y=abnormal|f(x_i)) = \frac{1}{1+ exp(wf(x_i)+b)}
\end{equation}
where $f(x_i)$ is the margin of sample $x_i$, $w$, $b$ are the parameters of the probablistic scaling model. The larger the probability is, the more likely the observed footstep is abnormal.


\vspace{20pt}
\noindent \uppercase{\textbf{Evaluation Through Real-World Walking Experiments}} \vspace{12pt} 

To evaluate the effectiveness of our approach, we conducted a real-world walking experiment with 8 human subjects (ages 18-40 years old). All experiments were conducted in accordance with the approved IRBs.

\textbf{Experiment Setup}

\begin{figure}[!t] 
  \centering{\includegraphics[width=\textwidth]{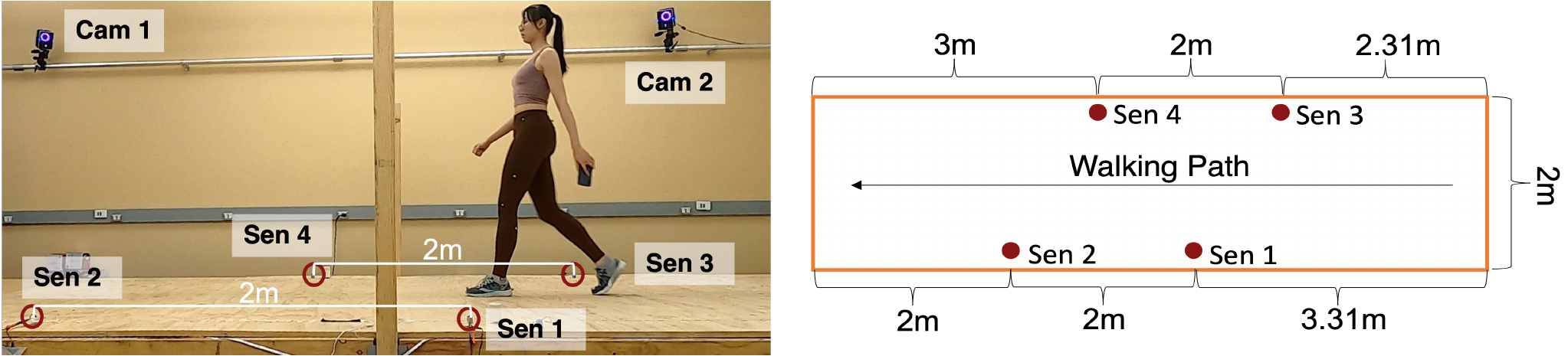}}
    \caption{Experiment setup with geophones (data) and Vicon cameras (ground truth).}
\label{fig:experiment}%
\end{figure}

The experiment setup involves four geophones for data collection and a Vicon Motion Capture system for ground truth. The geophones are placed at the side of a wooden walkway as shown in Figure~\ref{fig:experiment}. 

During the experiment, 5 test subjects walk across the walkway 30-40 times using their natural gait (which may or may not have abnormal contacts). 3 control subjects walk according to four types of abnormal gaits to simulate the abnormal contacts, including midfoot strike, toe-walking, dragging, and walking asymmetry.

\begin{figure}[!t] 
  \centering{\includegraphics[width=0.9\textwidth]{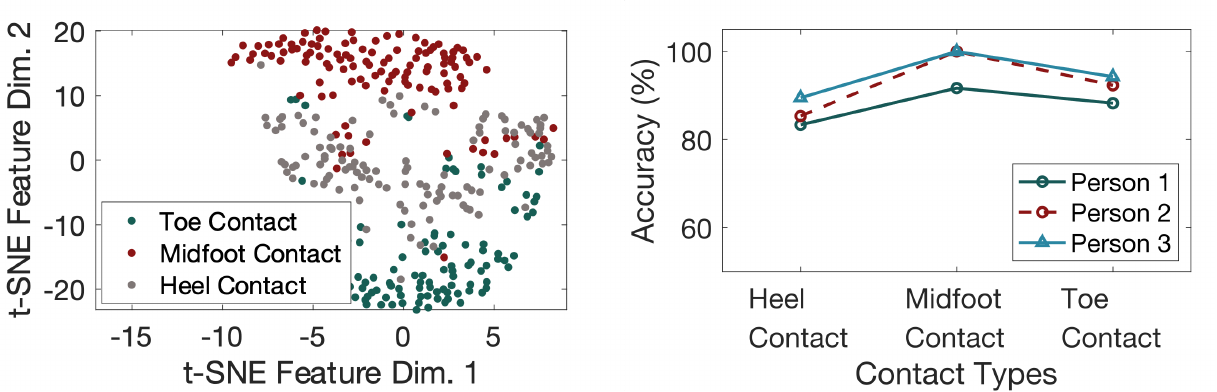}}
    \caption{Contact types visualized in time-frequency response features using t-SNE (left). Accuracy of contact type prediction for three subjects (right). }
\label{fig:acc_contact_type}%
\end{figure}

\textbf{Results and Discussion}

\textit{Contact Type Prediction.} Our approach has an average of 91.6\% and 96.7\% accuracy in predicting contact types and duration using the features extracted from the time-frequency response analysis. 
Figure~\ref{fig:acc_contact_type} (left) shows that these features produce distinct clusters for different contact types through t-SNE visualization. Figure~\ref{fig:acc_contact_type} (right) shows consistent accuracy of the 3 control subjects.

\begin{figure}[!t] 
  \centering{\includegraphics[width=0.9\textwidth]{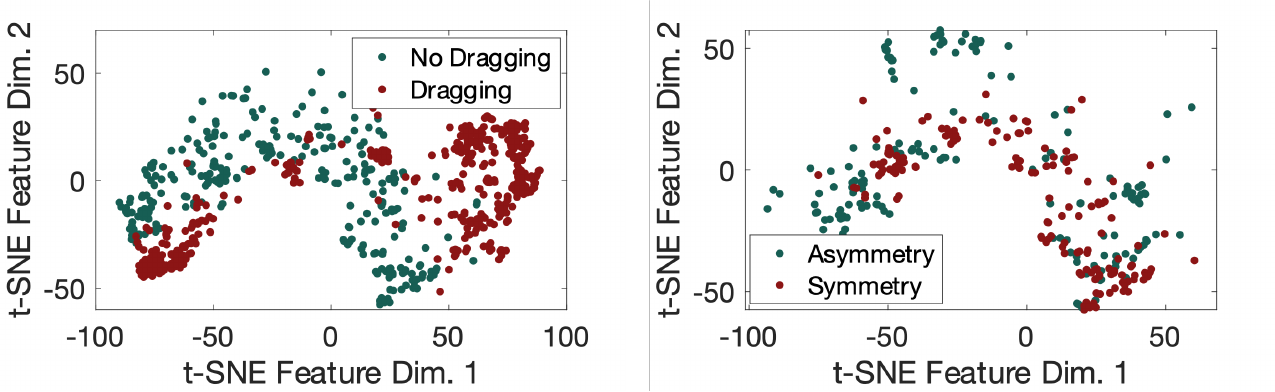}}
    \caption{Dragging and asymmetry visualized in time-frequency features using t-SNE.}
\label{fig:abnormaltsne}%
\end{figure}

\begin{figure}[!t] 
  \centering{\includegraphics[width=0.9\textwidth]{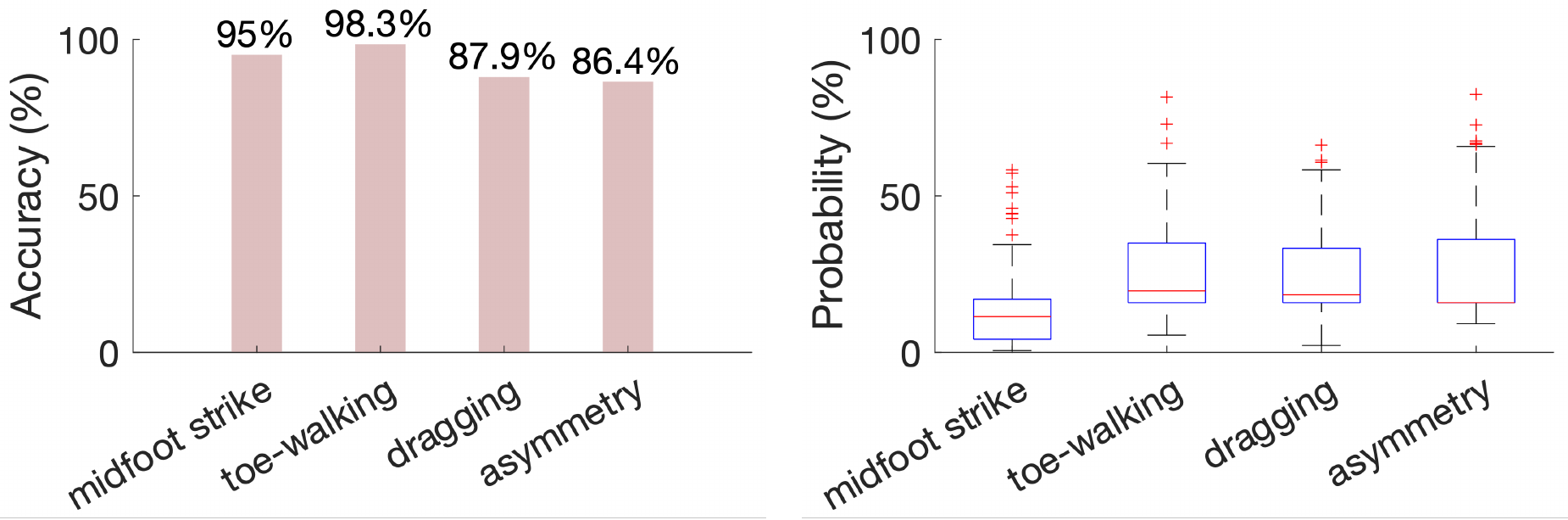}}
    \caption{Mean accuracy of abnormal gait detection among 8 participants (left). Probability of having abnormal contacts for a sample subject (right).}
\label{fig:abnormal}%
\end{figure}

\textit{Abnormal Gait Detection.} Our approach has an average of 91.9\% accuracy in detecting four types of abnormal contacts (see Figure~\ref{fig:abnormal} (left)). Figure~\ref{fig:abnormaltsne} shows the features from time-frequency analysis form distinct clusters for dragging and asymmetry gait, respectively.
Through our similarity-aware model, we compute the probability of having abnormal contacts for each test subject. Figure~\ref{fig:abnormal} (right) shows a sample subject's probability profile among the four abnormal types as an example. For this sample subject, the median values are relatively low despite the variation among footsteps, which means his/her gait is unlikely to be abnormal.

\vspace{20pt}

\noindent \uppercase{\textbf{CONCLUDING REMARKS}} \vspace{12pt}

In this study, we leverage footstep-induced floor vibrations to detect gait abnormalities in foot-floor contacts during walking. To overcome the challenge of the complex force transfer mechanism during the contacts, we characterize the vibration signals induced by three typical foot-floor contacts through dynamics formulation and controlled experiments. We introduce an abnormal gait detection framework that models contact time and type to make predictions on the abnormal contact probability. We evaluate our approach through a real-world experiment and results show promising accuracy in predicting contact types and detecting abnormal gait patterns.

\vspace{20pt}
\noindent \uppercase{\textbf{Acknowledgment}} \vspace{12pt}

This work was funded by the U.S. National Science Foundation (under grant number NSF-CMMI-2026699). The views and conclusions contained here are those of the authors and should not be interpreted as necessarily representing the official policies or endorsements, either express or implied, of any University, the National Science Foundation, the United States Government or any of its agencies.

\vspace{20pt}

\small 

\bibliographystyle{iwshm}
\bibliography{ref}


\end{document}